\documentclass[reprint,aps,pre,superscriptaddress]{revtex4-2}
\usepackage{amsmath}
\usepackage{amsthm}
\usepackage{amsfonts}
\usepackage{amssymb}
\usepackage{bm}
\usepackage{enumitem}
\usepackage{graphicx}
\usepackage{caption}
\usepackage{dsfont}
\graphicspath{{figures/}}
\usepackage{epstopdf}
\usepackage{hyperref}

\newcommand{\pderiv}[2]{\frac{\partial {#1}}{\partial {#2}}}

\let\originalleft\left
\let\originalright\right
\renewcommand{\left}{\mathopen{}\mathclose\bgroup\originalleft}
\renewcommand{\right}{\aftergroup\egroup\originalright}

\begin{document}
\title{ Theory of chemically driven pattern formation in phase-separating liquids and solids }

\author{Hongbo Zhao}
\affiliation{Department of Chemical Engineering, Massachusetts Institute of Technology \\ 77 Massachusetts Avenue, Cambridge, MA 02139}
\author{Martin Z. Bazant}
\affiliation{Department of Chemical Engineering, Massachusetts Institute of Technology \\ 77 Massachusetts Avenue, Cambridge, MA 02139}
\affiliation{Department of Mathematics, Massachusetts Institute of Technology \\ 77 Massachusetts Avenue, Cambridge, MA 02139}
\date{\today}

\begin{abstract}
  Motivated by recent experimental and theoretical work on the control of phase separation by (electro-)autocatalytic reactions, we analyze pattern formation in externally driven phase separating systems described by a generalization of the Cahn-Hilliard and Allen-Cahn equations combining nonlinear reaction kinetics with diffusive transport. The theory predicts that phase separation can be suppressed by driven autoinhibitory reactions when chemically driven at a sufficiently high reaction rate and low diffusivity, while autocatalytic reactions enhance phase separation. Analytical stability criteria for predicting the critical condition of suppressed phase separation based on linear stability analysis track the history dependence of pattern formation and agree well with numerical simulations. By including chemo-mechanical coupling in the model, we extend the theory to solids, where coherency strain alters the morphology and dynamics of driven phase separation. We apply this model to lithium iron phosphate nanoparticles and simulate their rate-dependent electrochemical charging and discharging patterns, paving the way for a quantitative understanding of the effect of reaction kinetics, diffusion, and mechanics on the electrochemical performance of energy materials. The theory may also find applications to microstructure formation in hardening cement paste, as well as membraneless organelle formation in biological cells by chemically controlled liquid-liquid phase separation.
\end{abstract}

\maketitle

\section{Introduction} \label{sec::intro}
When a phase-separating mixture is in a thermodynamically unstable state -- the free energy $G$ is concave with respect to certain variations in the concentration of the constitutive chemical species ($\delta^2 G<0$), the system can find a lower energy state by spontaneously separating into phases with different chemical compositions.
The dynamics of this process, known as spinodal decomposition \cite{Cahn1961,Bray2002}, namely the formation of interspersed domains of different phases and their coarsening over time as the free energy of the system continues to decrease by the reduction in its interfacial energy between the phases, has been studied extensively through mean-field theory\cite{Lifshitz1961,Siggia1979,Bray2002,Binder1987,Valls1993,Furukawa2000,Koga1991}, molecular dynamics simulations\cite{Velasco1996,Velasco1993,Laradji1996} and experiments\cite{Nishi1975,Huang1974,Hashimoto1988}.

A well-known mean-field model that describes spinodal decomposition in systems that are dominated by molecular diffusion and satisfy mass conservation is the Cahn-Hilliard equation \cite{CahnJohnE.1958}, in which the diffusive flux driven by the gradient of the chemical potential leads to uphill diffusion in the thermodynamically unstable region. On the other hand, the Allen-Cahn equation (A-H) \cite{Kardar2007,Allen1972} and various stochastic pattern-forming models of Hohenberg and Halperin~\cite{hohenberg1977theory} have been used to describe phase-transforming systems with nonconserved order parameters, whose dynamics follow the direction of gradient descent of the free energy. 

Building on the theory of non-reactive phase separation, there has been a growing interest in the study of phase separation in chemically driven systems, due to its application in a wide range of fields, such as externally driven electrochemical reactions in phase-separating energy materials \cite{Bazant2013,Bazant2017,Bai2011,Lim2016}, ATP-driven reactions and cell proliferation related to phase separation in biology\cite{Zwicker2015,Zwicker2017,Berry2018,Zwicker2022,Cates2010}, and reactive colloidal cluster formation in hardening cement paste~\cite{petersen2018phase,ioannidou2016mesoscale} and nanoparticle aggregation and gelation~\cite{weitz1984fractal,zaccarelli2008gelation}.
As a result of chemical reactions, the system deviates from the downhill trajectory toward free energy minima. In many works coupling Cahn-Hilliard equation with reactions that break detailed balance, emergent patterns such as suppression of Ostwald ripening, and even dynamic morphologies that resemble protocell division\cite{Glotzer1995,Cates2010,Zwicker2015,Zwicker2017} have been observed.

In recent years, a general thermodynamically consistent model of chemical kinetics in phase separating systems has been developed that generalizes Cahn-Hilliard and Allen-Cahn equations for nonlinear reaction kinetics coupled with diffusion-driven phase separation ~\cite{Bazant2013}. The model has led to a general theory of thermodynamic stability for driven open systems far from equilibrium, which predicts suppression of phase separation, as well as pattern formation in stable mixtures, driven by chemical reactions~\cite{Bazant2017}. The theory elucidates reaction-driven phase transformations in electrochemistry, especially in lithium-ion batteries where many widely used lithium intercalation materials such as lithium iron phosphate (LFP) nanoparticles and graphite undergo phase transformation \cite{Lim2016,Bai2011,Cogswell2012,Cogswell2013,Nadkarni2018,Fraggedakis2020a,Cogswell2018a,Zhao2023LFP,Gao2021,Guo2016}, along with applications in electrodeposition, adsorption, and hydrogen storage\cite{Horstmann2013,Seri-Levy1993,Baldi2014,Narayan2017}. The stability condition of the model predicts that autocatalytic or autoinhibitory reactions, which satisfy detailed balance, can strongly compete with phase separation, thereby controlling the thermodynamic stability of reactive mixtures. Indeed, experiments have confirmed the predicted role of electro-autocatalysis in the suppression of phase separation, dynamic surface phases, and population dynamics in LFP\cite{Lim2016,Li2018a,Li2014,Zhao2023LFP,Koo2023,Zhao2019,Smith2017}, as well as the enhancement of phase separation in lithium layered oxides \cite{Park2021}.

While the theory has found wide applications in elucidating the phase behavior of electrochemical materials under charge and discharge, a quantitative analysis of how the coupling between reaction, diffusion, and (in some cases) solid mechanics affects pattern formation in these systems is still lacking. 
In the case of Li-ion batteries, there is growing experimental evidence from {\it operando} x-ray microscopy that all three of these determine phase-separation patterns in these materials \cite{lu20203d,cocco2013three,Ebner2013,Gent2016,Lim2016,Lin2017,Li2018a,cao2020emerging,Merryweather2021,Mefford2021,Gao2021,Xu2022}.  We are particularly motivated by the recent demonstration of learning reaction kinetics of lithium intercalation in LFP directly from images of driven pattern formation, which establish the validity of the model with unprecedented accuracy down to the nanoscale~\cite{Zhao2023LFP}.

In this article, we extend the original analysis of the general stability theory~\cite{Bazant2017} to include two-dimensional diffusion and chemo-mechanical coupling and compare the resulting analytical predictions with numerical simulations of the full model.
Based on the theory of reactive mixtures that we introduce in sec.~\ref{sec::model}, we aim to provide a quantitative prediction for the condition of phase separation for the general case of a reactive mixture detailed in sec.~\ref{sec::linear_stability} and \ref{sec::reaction_kinetics} and compare with numerical simulations at different reaction rates and diffusivities in sec.~\ref{sec::simulation}. 
Since the insertion of lithium into LFP is accompanied by a deformation of the crystal lattice, we then extend the theory to account for chemo-mechanical coupling in sec.~\ref{sec::chemomechanics} and apply it to LFP nanoparticle to simulate rate-dependent electrochemical charging and discharging patterns that are consistent with experiments.

\section{Model.} \label{sec::model}
\begin{figure}
    \centering
    \includegraphics[width=0.8\columnwidth]{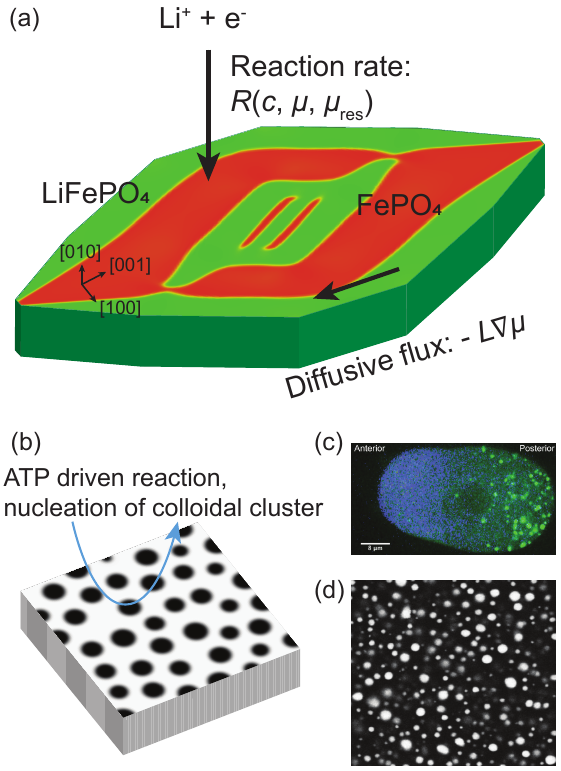}
    \caption{Chemically driven phase-separating systems. (a) Schematics of lithium intercalation in LFP with a superimposed simulation result. Lithium ions in the electrolyte are inserted into the crystal via the electrochemical reaction $\text{Li}^+ + e^{-1} + \text{FePO}_4 \rightarrow \text{LiFePO}_4$ at the (010) surfaces. The reaction rate $R$ is dependent on the concentration $c$ and chemical potential $\mu$ of lithium in LFP and the chemical potential of the reservoir species which includes $\text{Li}^+$ and electrons, hence $\mu_\text{res} = \mu_\text{Li}^+ - e\Delta \phi$, where $\Delta \phi$ is the interfacial voltage between the electrolyte and LFP. Lithium may also diffuse in the [100] and [001] directions on the surface or the bulk, and the diffusive flux is proportional to the gradient of the chemical potential. When the reaction is slow and diffusion is fast, LFP undergoes phase separation, forming lithium-rich (green) and lithium-poor (red) phases, with preferred interfacial orientation in the [101] direction due to anisotropic lattice deformation induced by lithium insertion. (b) Schematics of chemically reactive liquid-liquid phase separation driven by ATP reactions or nucleation of colloidal clusters. (c) C. elegans embryo maintains a gradient of MEX-5 protein and phase-separated P granules along the anterior-posterior axis \cite{Saha2016} (adapted with permission). (d) Phase separation in milk protein \cite{deBont2002} (adapted with permission).}
    \label{fig::schematics}
\end{figure}

In this work, we study \emph{phase-separating, chemically driven, and open systems}, where the phase separating chemical species M is not conserved and can be added or removed from the system with the chemical in the reservoir $\text{M}_\text{res}$ through a bulk reaction: $\text{M}_\text{res}\to \text{M}$, that is, the exchange between the system and reservoir occurs everywhere in the bulk\cite{Bazant2013}. This situation describes, for example, adsorption or electrochemical intercalation in thin platelet particles \cite{Singh2008,Bai2011,Cogswell2012,Bazant2017} where the depth in the direction normal to the surface is sufficiently small or the chemical composition in the depth direction is homogeneous, hence the entire system effectively undergoes bulk reaction through the reactive surface that is exposed to the chemical reservoir.
An example is shown in Fig.~\ref{fig::schematics}(a), where lithium ions in the electrolyte participate in the electrochemical intercalation reaction at the (001) surfaces into thin platelet $\text{FePO}_4$ particles and form $\text{LiFePO}_4$, driven by the electrical voltage applied across the solid-electrolyte interface. More broadly, as illustrated in Fig.~\ref{fig::schematics}(b), the model we consider is also generalizable to other reactive phase separating systems such as membraneless organelles in living cells that are constantly driven out of equilibrium by the intracellular reactions (Fig.~\ref{fig::schematics}c), and aggregation of colloidal particles (Fig.~\ref{fig::schematics}(d)).

The reaction rate $R$ is dependent on the local concentration $c$ and chemical potential $\mu$ of M in the system, and the chemical potential of $\text{M}_\text{res}$ in the reservoir $\mu_\text{res}$. The driving force or the chemical affinity for the reaction is $\mu_\text{res}-\mu$, which can be controlled by tuning the reservoir chemical potential, or via electrical voltage in the case of electrochemical reactions. The reaction rate is zero under zero driving force ($\mu=\mu_\text{res}$), hence satisfying detailed balance. In addition, a positive chemical affinity corresponds to a positive reaction rate, that is, $(\mu_\text{res}-\mu)\cdot R>0$, hence the reaction satisfies the condition of positive entropy production \cite{Bazant2017}. We will study specific forms of reaction kinetics and their influence on phase separation in sec.~\ref{sec::reaction_kinetics}.

In addition to bulk reaction, the species M can also diffuse in the system, driven by the gradient in the chemical potential $\mu$. For example, in the case of adsorption, the adsorbed molecules may freely diffuse on the 2D surface, and in LFP nanoparticles, lithium may also diffuse laterally in the bulk or on the surface \cite{Li2018a,malik2010}, as indicated in Fig.~\ref{fig::schematics}.
Therefore, the dynamics of the concentration of M is governed by the following reaction-diffusion equation \cite{Bazant2013,Bazant2017},
\begin{equation} \label{eqn::PDE}
  \pderiv{c}{t} = \nabla \cdot (L \nabla \mu) + R(c,\mu,\mu_\text{res}),
\end{equation}
where $t$ is time, $L$ is the Onsager coefficient relating diffusive flux to the chemical potential gradient, and $c$ is nondimensionalized by the maximum concentration so that $c\in[0,1]$.

Without reaction, the equation reduces to the Cahn-Hilliard equation, and without diffusion, the equation is the Allen-Cahn-reaction (ACR) equation \cite{Bazant2013}, generalizing the Allen-Cahn equation and various related models in statistical physics~\cite{hohenberg1977theory} to externally driven chemical reactions.
Following the Cahn-Hilliard formulation, the free energy of a nonuniform or phase-separating system $G$ is given by
\begin{equation}
  \label{eqn::chemical_free_energy}
  G[c] = \int{\left( g_h(c) + \frac{1}{2} \kappa |\nabla c|^2 \right) d\mathbf{x}}.
\end{equation}
where $g_h(c)$ is the homogeneous free energy density that is a function of the local concentration $c$. The second term is the energy associated with the concentration gradient that results in a diffuse interface between phases at equilibrium. $\kappa$ is the gradient energy coefficient.
The chemical potential $\mu$ is defined to be the variational derivative of the free energy functional $G[c]$, $\mu \equiv \delta G / \delta c$.
Hence
\begin{equation} \label{eqn::chemical_potential}
  \mu = \mu_h(c) - \kappa \nabla^2c,
\end{equation}
where $\mu_h(c) = \partial g_h / \partial c$ is the homogeneous chemical potential.
For phase separation to occur, the homogeneous free energy density $g_h(c)$ has a double-well shape. At equilibrium, the two phases have the same chemical potential $\mu_0$, and their concentrations $c_1$ and $c_2$ correspond to the energy minima obtained by common tangent construction -- $g_h(c_2)-g_h(c_1) = \mu_0 (c_2-c_1)$ for \cite{CahnJohnE.1958}. With a homogeneous concentration $c(\mathbf{x})=c_0$, the second order variation of free energy $\delta^2 G$ becomes negative when $\mu_h'(c_0)<0$, indicating that the system is thermodynamically unstable and spinodal decomposition occurs spontaneously. The spinodal region is defined as the region where $\mu_h'(c_0)<0$.

\section{Linear stability analysis}
\label{sec::linear_stability}
In the following sections, for convenience, we define the following normalized quantities. We normalize all chemical potentials ($\mu$,$\mu_\text{res}$, and $\mu_h$) by thermal energy denoted by a tilde, for example,  $\tilde{\mu} = \mu/(k_B T)$, where $k_B$ is Boltzmann constant and $T$ is temperature. Additionally, we define $\tilde{\kappa}=\kappa / (k_B T)$, and $\tilde{L} = L k_B T$.
By perturbing Eqs.~\ref{eqn::PDE} and \ref{eqn::chemical_potential} around a uniform solution $c(\mathbf{x})=c_0$ with Fourier modes $\delta \hat{c} e^{\sigma t + \mathbf{k}\cdot \mathbf{x}}$, we obtain the dispersion relation~\cite{Bazant2017}.
\begin{equation} \label{eqn::dispersion}
  \sigma = \pderiv{R}{c} + \left( \pderiv{R}{\tilde{\mu}} - \tilde{L} k^2 \right) \left( \tilde{\mu}_h' + \tilde{\kappa} k^2 \right),
\end{equation}
where all the derivatives are taken at $c=c_0$ and $k=|\mathbf{k}|$.

When there is no reaction ($R=0$), Eq.~\ref{eqn::PDE} reduces to Cahn-Hilliard equation \cite{CahnJohnE.1958}. Accordingly, Eq.~\ref{eqn::dispersion} indicates that when the system is thermodynamically unstable, that is, the free energy is nonconvex $\tilde{\mu}_h'<0$, also known as the spinodal region for a phase separating system, and when it is purely diffusive, the system undergoes spinodal decomposition, forming patterns at wavenumber $k_\text{sp} = \sqrt{-\tilde{\mu}_h' / 2\tilde{\kappa}}$, which corresponds to the maximum instability growth rate $\sigma_\text{sp} = \sigma(k_\text{sp};R=0) = L\tilde{\mu}_h^{\prime2}/4\tilde{\kappa}$.

When the system is reactive, the effect of reaction on phase separation is reflected in Eq.~\ref{eqn::dispersion} in two terms. The first term is $\partial R / \partial c$, which we call solo-autocatalytic rate\cite{Bazant2017} since it reflects the explicit dependence of the reaction rate on species concentration (rather than through the dependence of $R$ on $\tilde{\mu}$, which also depends on $c$). The second reaction-related term in Eq.~\ref{eqn::dispersion} involves $\partial R / \partial \tilde{\mu}$. Typically, the rate of reaction $\text{M}_\text{res}\to \text{M}$ increases with increasing reaction affinity ($\tilde{\mu}_\text{res}-\tilde{\mu}$), therefore, $\partial R/\partial \tilde{\mu}<0$.
However, we note that in electrochemistry, it is possible to have $\partial R/\partial \tilde{\mu}>0$ in Marcus kinetics for outer-sphere electron transfer\cite{Marcus1993}, which can have a destabilizing effect and promote phase separation\cite{Bazant2013}. 

Following Ref.~\cite{Bazant2017}, we define Damkholer number to be the ratio of the differential reaction rate $\partial_{\tilde{\mu}} R \cdot \tilde{\mu}_h'$ to the rate of spinodal decomposition,
\begin{equation}
  \text{Da} \equiv \frac{\tilde{\kappa} \partial_{\tilde{\mu}} R}{\tilde{L} \tilde{\mu}_h'}.
\end{equation}
When $0 \leq \text{Da}<1$, that is, diffusion is fast compared to reaction and the system is in the spinodal region ($\tilde{\mu}_h'<0$), the most unstable wavenumber is
\begin{equation}
  k_\text{max} \equiv \text{argmax}_{k} \sigma(k) = k_\text{sp} \sqrt{1-\text{Da}},
\end{equation}
and the corresponding instability growth rate is
\begin{equation}
  \sigma_\text{max} = \pderiv{R}{c} + \pderiv{R}{\tilde{\mu}}\tilde{\mu}_h' \frac{(1+\text{Da})^2}{4\text{Da}}.
\end{equation}
When $\text{Da} \geq 1$ (diffusion is slow compared to reaction and $\tilde{\mu}_h'<0$), or $\text{Da}<0$ (the system is outside the spinodal region, $\tilde{\mu}_h'>0$),
the most unstable wavenumber is $k_\text{max}=0$, and $\sigma_\text{max} = \partial_c R + \tilde{\mu}_h' \partial_{\tilde{\mu}} R$.

In summary, we can unify all cases above by defining the following function
\begin{equation}
  F(\text{Da}) = \frac{(1+ \text{max} \{ \text{Da}^{-1},1\} )^2}{4 \text{max} \{ \text{Da}^{-1},1 \} }.
\end{equation}
Hence the maximum instability growth rate is
\begin{equation} \label{eqn::sigma_max}
  \sigma_\text{max} = \pderiv{R}{c} + \pderiv{R}{\tilde{\mu}} \tilde{\mu}_h' F(\text{Da}),
\end{equation}
and the most unstable wavenumber is
\begin{equation} \label{eqn::k_max}
  k_\text{max} = k_\text{sp} \sqrt{1-\text{max}\{\text{Da}^{-1},1\}^{-1}}.
\end{equation}

In the spinodal region and in the limit of fast diffusion ($0<\text{Da} \ll 1$), $F(\text{Da}) \sim 1/(4\text{Da})$,
\begin{equation} \label{eqn::fast_diffusion}
  \sigma_\text{max} \approx \pderiv{R}{c} + \frac{L \tilde{\mu}_h^{\prime 2}}{4\tilde{\kappa}}.
\end{equation}

As mentioned above, when the system is purely diffusive, phase separation occurs spontaneously in the spinodal region. However, Eq.~\ref{eqn::sigma_max} suggests that, depending on $\partial_c R$, $\partial_{\tilde{\mu}} R$, and $\text{Da}$, reactive systems can become stablized against perturbations ($\sigma_\text{max}<0$). 
Therefore, we now discuss the condition for the suppression of phase separation at any given $c_0$ using the dispersion relation and later in section \ref{sec::simulation} compare with numerical simulations.

When the system is at the base state $c(\mathbf{x})=c_0$, the sign of $\sigma_\text{max}$ whether perturbation introduced as this moment will grow in time. Therefore, the stability condition in the short time limit is $\sigma_\text{max}<0$.

However, as explained in sec.~\ref{sec::intro}, we are particularly interested in the stability of the system when it is chemically driven at a nonzero rate. For example, in electrochemistry, the reaction rate of Faradaic reactions can be specified by controlling the current\cite{Newman2012}, while the interfacial voltage is adjusted accordingly (which corresponds to varying $\mu_\text{res}$) such that the average reaction rate is at a specified value $\bar{R}$, that is $\bar{R} = \int{\partial_t c d\mathbf{x}}/\int{d\mathbf{x}}$.
For a uniform base state, the reaction rate $R$ is uniform and hence equal to $\bar{R}$. For intercalation and adsorption, there exists a maximum concentration and the system will eventually reach the fully reacted state. Recall that $c$ is normalized such that $c_0=1$ corresponds to the fully reacted state. Therefore, instability can only develop over a time scale of $1/R$. 
Following Ref. \cite{Bai2011}, we compare the instability growth rate $\sigma_\text{max}$ with the rate of filling the system with M from $c_0=0$ to $c_0=1$ (or the other direction) if the reaction takes place at a constant rate $R$, and use $\sigma_\text{max}/|R|<1$ as the stability criterion.


However, the condition above does not take into account the history of the temporal evolution of $c_0$. As the reaction proceeds, the average fraction $c_0$ changes and so does $\sigma_\text{max}$. Hence, the past trajectory $c_0(t)$ for $t<t_0$ determines whether phase separation occurs at $t=t_0$.
Therefore, we argue that the stability criterion can be further refined by considering the dependence of $\sigma$ on $c_0$.
The perturbation $\delta \hat{c}$ grows according to $\partial \delta \hat{c} /\partial t = \sigma(k;t) \delta \hat{c}$, where the dependence of $\sigma$ on $t$ is through $c_0(t)$ and $\mu_\text{res}(t)$.
Therefore, for small perturbations, $\delta \hat{c}(t) = \delta \hat{c}(0)\exp{\int_0^t{\sigma dt}}$.
As an upper bound for the growth of perturbation, we define instability amplification factor $K=\int_0^t{\sigma_\text{max} dt}$, which is the integral of the maximum instability growth rate over time. For a purely reactive and spatially non-interacting system, Ref. \cite{Zhao2020} has shown that with small perturbations the variance of $c$ grows as $e^{2K}$ in time, and provided an estimate for the stability condition, which we extend in our study to reactive-diffusive systems: $0 < c_0(t=0) \pm \sigma_0 e^K < 1$, where $\sigma_0$ is the standard deviation of the initial perturbation, in other words, the perturbation does not go beyond $[0,1]$.

Another equivalent definition of $K$ is $K = \int_{c_0(0)}^{c_0(t)}{\sigma_\text{max}/R dc_0}$,
where the normalized maximum instability growth rate is
\begin{equation} \label{eqn::normalized_sigma_general}
  \frac{\sigma_\text{max}}{R} =  \pderiv{\ln{R}}{c} + \pderiv{\ln{R}}{\tilde{\mu}} \tilde{\mu}_h' F(\text{Da}).
\end{equation}

In sec. \ref{sec::simulation} and \ref{sec::chemomechanics}, we will further elaborate on the following three stability criteria by comparing with simulations: (A) $\sigma_\text{max}<0$; (B) $\sigma_\text{max}/|R|<1$; (C) $0 < c_0(t=0) \pm \sigma_0 e^K < 1$, will 

\section{Reaction kinetics}
\label{sec::reaction_kinetics}
In this section, we consider specific forms of reaction kinetics $R$.
Starting from transition state theory, one can derive a commonly used form of reaction kinetics \cite{Bazant2013} which has a separable dependence on $c$, $\tilde{\mu}$ and the reaction affinity $\Delta \tilde{\mu} \equiv \mu_\text{res}-\tilde{\mu}$,
\begin{equation} \label{eqn::R_c_Delta_mu}
  R = R_0(c,\tilde{\mu}) g(\Delta \tilde{\mu}).
\end{equation}
where $R_0$ is a positive kinetic prefactor, and $g(\Delta \tilde{\mu})$ is the functional form of the dependence on affinity. The reaction kinetics satisfies detailed balance, that is, $g(0)=0$, and positive entropy production rate $g(\Delta \tilde{\mu}) \cdot \Delta \tilde{\mu}>0$.
At low affinity when the system is close to equilibrium, linear irreversible thermodynamics applies and $g(\Delta\tilde{\mu}) \approx \Delta \tilde{\mu}$\cite{Kondepudi2014}.

Based on Eq.~\ref{eqn::normalized_sigma_general}, for reaction kinetics that follows Eq.~\ref{eqn::R_c_Delta_mu}, the normalized maximum instability growth rate is
\begin{equation} \label{eqn::normalized_sigma}
  \frac{\sigma_\text{max}}{R} =  \pderiv{\ln{R_0}}{c} + \left( \pderiv{\ln{R_0}}{\tilde{\mu}} - \frac{g'}{g}\right) \tilde{\mu}_h' F(\text{Da}).
\end{equation}

Butler-Volmer kinetics, which is often used in electrochemistry, belongs to this category \cite{Newman2012}:
\begin{equation} \label{eqn::BV}
  g(\Delta \tilde{\mu}) = e^{\alpha \Delta \tilde{\mu}} - e^{-(1-\alpha) \Delta \tilde{\mu}},
\end{equation}
where $\alpha$ is the symmetry factor and $0<\alpha<1$.

When the system is close to equilibrium ($\Delta \tilde{\mu} \to 0$), $g\to0$, the second term in Eq.~\ref{eqn::normalized_sigma} dominates. In the spinodal region, $\sigma_\text{max}>0$, the system is unstable.

Within the spinodal region, if one drives the system further away from equilibrium, and if the prefactor $R_0$ depends only on $c$ ($\partial_\mu R_0=0$), $\sigma_\text{max}/|R|$ decreases with increasing magnitude of reaction rate $|R|$ (see Appendix \ref{appendix::norm_instability} for proof), and at infinitely large reaction rates,
\begin{align}
  \lim_{R\to\infty}{\frac{\sigma_\text{max}}{|R|}} &= \pderiv{\ln{R_0}}{c} - \alpha \tilde{\mu}_h'F(\text{Da})  \label{eqn::normalized_sigma_BV_plus_infty} \\
  \lim_{R\to-\infty}{\frac{\sigma_\text{max}}{|R|}} &= -\pderiv{\ln{R_0}}{c} - (1-\alpha) \tilde{\mu}_h'F(\text{Da})   \label{eqn::normalized_sigma_BV_minus_infty}.
\end{align}
Because $\tilde{\mu}_h'<0$ in the spinodal region, the second term is always positive and destabilizing. However, the reaction can alter the stability via the first term. Therefore, it is possible to kinetically stabilize a thermodynamically unstable system at a high enough reaction rate $|R|$ by having a concentration-dependent prefactor $R_0' \neq 0$ (a sufficient condition is $|\partial_c \ln{R_0}|> -\tilde{\mu}_h'$), although it is only possible to make $\sigma_\text{max}<0$ for either forward or backward reaction.

Eqs.~\ref{eqn::normalized_sigma_BV_plus_infty} and \ref{eqn::normalized_sigma_BV_minus_infty} show that the asymmetry of the instability between forward and backward reactions can come from the solo-autocatalysis and the asymmetric dependence on affinity (when $\alpha\neq 0.5$).



More generally, one can derive a reaction rate based on transition state theory.
Ref. \cite{Bazant2013} derived generalized Butler-Volmer kinetics where the prefactor $R_0$ depends explicitly on both $c$ and $\mu$
\begin{equation} \label{eqn::generalized_R0}
  R_0(c,\mu) = \frac{k_0 e^{\alpha\tilde{\mu}}}{\gamma_\dagger(c)},
\end{equation}
where $\gamma_\dagger$ is the activity coefficient of the transition state which depends on $c$ in concentrated solutions.
Using this model, we have $\partial_c\ln{R_0} = - (\ln{\gamma_\dagger})'$, and with $g(\Delta \tilde{\mu})$ from Eq.~\ref{eqn::BV},
\begin{equation}
  \pderiv{\ln{R}}{\tilde{\mu}} = \frac{1}{1-e^{\Delta \tilde{\mu}}}.
\end{equation}
Therefore, based on Eq.~\ref{eqn::normalized_sigma_general}, it follows that the generalized Butler-Volmer kinetics (Eq.~\ref{eqn::generalized_R0} and \ref{eqn::BV}) also has the property that $\sigma_\text{max}/|R|$ decreases with increasing $|\Delta \tilde{\mu}|$ (or increasing $|R|$). The minimum is achieved at infinitely large $|R|$,
\begin{align}
  \lim_{R\to\infty}{\frac{\sigma_\text{max}}{|R|}} &= (\ln{\gamma_\dagger})' \label{eqn::normalized_sigma_2_plus_infty}\\
  \lim_{R\to-\infty}{\frac{\sigma_\text{max}}{|R|}} &= -(\ln{\gamma_\dagger})' - \tilde{\mu}_h' F(\text{Da}) \label{eqn::normalized_sigma_2_minus_infty},
\end{align}
which shows that under the generalized Butler-Volmer kinetics, it is possible to make $\sigma_\text{max}<0$ via reaction at high enough reaction rates, although in the spinodal region, it is only possible to stabilize either forward or backward reaction. Eq.~\ref{eqn::normalized_sigma_2_plus_infty} shows that at infinitely fast and positive reaction rates, the stability solely depends on $\gamma_\dagger$, and is independent of the thermodynamics.

In summary, we come to the conclusion that for Butler-Volmer kinetics with a prefactor that does not depend on $\mu$ or a prefactor that follows Eq.~\ref{eqn::generalized_R0}, the system may be stabilized at a high enough reaction rate.
The critical reaction rate $R_\text{crit}$ (or affinity)  above which phase separation is suppressed can be solved using Eq.~\ref{eqn::normalized_sigma} via the stability criteria discussed in sec.~\ref{sec::linear_stability}.

Following Eq.~\ref{eqn::normalized_sigma_general}, the instability amplification factor $K$ can be written as
\begin{equation}
  K = \left. \ln{R_0} \right|_{c_0(0)}^{c_0(t)} - \int_{c_0(0)}^{c_0(t)}{\pderiv{\ln{R}}{\tilde{\mu}} \tilde{\mu}_h' F(\text{Da}) dc_0}.
\end{equation}

When $\alpha=0.5$, Eq.~\ref{eqn::normalized_sigma} can be expressed explicitly in terms of $R$ and eliminating $\Delta \tilde{\mu}$ (see Appendix Eqs.~\ref{eqn::normalized_sigma_BV} and \ref{eqn::normalized_sigma_generalized_BV}).


To study how the critical reaction rate to suppress phase separation varies with the ratio of reaction to diffusion rate, we define a nominal Damkholer number that does not depend on $c$, $\mu$ and $\mu_\text{res}$,
\begin{equation}
  \text{Da}_0 \equiv \frac{\tilde{\kappa} k_0}{\tilde{L}_0},
\end{equation}
where $k_0$ is a characteristic kinetic constant that appears in $R_0$; $\tilde{L}_0$ is a characteristic mobility (since the Onsager coefficient $\tilde{L}$ may also depend on $c$).

In the limit of fast diffusion ($0<\text{Da} \ll 1$), for reaction kinetics that follows the separable form Eq.~\ref{eqn::R_c_Delta_mu}, Eq.~\ref{eqn::fast_diffusion} can be written as
\begin{equation} \label{eqn::sigma_Da_ll_1}
  \frac{\sigma_\text{max}}{R} \sim  \frac{R_0'}{R_0} +  \frac{ 1}{4\text{Da}_0} \frac{k_0\tilde{L} \tilde{\mu}_h^{\prime2}}{\tilde{L}_0 R}.
\end{equation}
From this, we see that whichever stability criterion is chosen, the critical reaction rate (if it exists) scales as $R_\text{crit}/k_0 \propto \text{Da}_0^{-1}$ in the fast-diffusion limit.
On the other hand, when $\text{Da}>1$, $F(\text{Da})=1$, in the fast-reaction regime, the critical reaction rate is independent of $\text{Da}_0$.

%
%

\section{Simulations of a generic chemically driven phase-separating mixture} \label{sec::simulation}

\begin{figure*}[htb]
  \includegraphics[width=\textwidth]{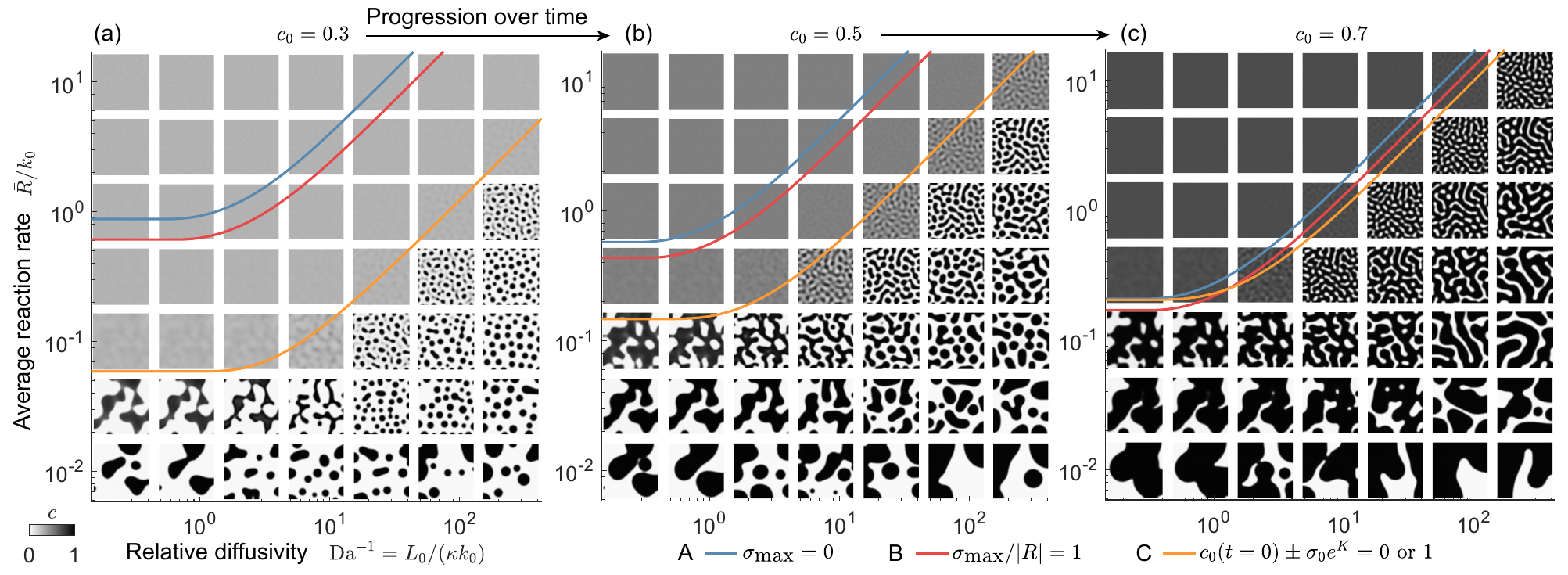}
  \caption{Pattern formation of a reactive-diffusive and phase-separating system in the $\text{Da}_0^{-1}-\bar{R}/k_0$ phase diagram. Simulations start from a uniform concentration field with added noise. Constant average reaction rate $\bar{R}$ is imposed. From left to right, (a-c) show snapshots when the average concentrations are 0.3, 0.5, and 0.7, respectively. All three panels are plotted on the same axis, hence some axis labels are omitted. The colored solid curves denote the boundaries of the analytical stability criteria: (A) $\sigma_\text{max}<0$; (B) $\sigma_\text{max}/|R|<1$ and (C) $0 < c_0(t=0) \pm \sigma_0 e^K < 1$. The predicted region of stability is above the solid curves.}
  \label{fig::no_stress}
\end{figure*}

In this section, we study a generic chemically driven phase-separating binary mixture whose free energy is modeled by the regular solution model, which has been used in various physical systems ranging from binary alloys and lithium intercalation materials to biological liquid-liquid phase separation \cite{CahnJohnE.1958,Bray2002,Bazant2013,Berry2018}. The mixture can contain two different chemical species or one chemical species and vacancies. The free energy consists of the ideal entropy of mixing and the enthalpy of mixing, which is proportional to the product of the concentrations under the mean-field approximation \cite{CahnJohnE.1958},
\begin{equation} \label{eqn::regular_solution}
  \frac{g_h(c)}{k_B T} = c\log{c} + (1-c)\log{(1-c)} + \Omega c(1-c),
\end{equation}
When the interaction parameter $\Omega>2$, the free energy is concave in the spinodal region $[c_-,c_+]$ where $c_\pm = (1\pm \sqrt{1-2/\Omega})/2$. In this section, we set $\Omega = 4$.

The reaction kinetics follows Eqs.~\ref{eqn::R_c_Delta_mu} and \ref{eqn::generalized_R0} with symmetric dependence on the affinity ($\alpha=0.5$). We assume that the transition state of the reaction $\text{M}_\text{res}\to \text{M}$ requires the presence of vacancies, hence $\gamma_\dagger = 1-c$ \cite{Bazant2013}. Since the diffusive flux is proportional to the chemical potential gradient multiplied by the concentration of M, and that diffusion of M also requires the presence of vacancy\cite{Bazant2013}, we use $\tilde{L} = \tilde{L}_0 c(1-c)$.

We perform numerical simulation on a 2D rectangular domain of size $[50\sqrt{\tilde{\kappa}},50\sqrt{\tilde{\kappa}}]$ starting from a uniform concentration field $c_0(t=0)=0.1$ with additive white noise ($\sigma_0=0.02$) as the initial condition. A constant average reaction rate $\bar{R}$ is imposed while $\mu_\text{res}$ adjusts accordingly in time to satisfy this constraint. We impose no-flux ($\mathbf{n} \cdot \nabla \mu=0$) and non-wetting ($\mathbf{n} \cdot \nabla c=0$)\cite{CahnJohnE.1958} boundary conditions. The domain is triangulated with a maximum mesh edge length of $0.5\sqrt{\tilde{\kappa}}$ and Eq.~\ref{eqn::PDE} is solved using finite element method.

We calculate the critical reaction rate at any $\text{Da}_0$ based on the stability criteria discussed in section \ref{sec::linear_stability}. Because $\gamma_\dagger'<0$ and based on Eq.~\ref{eqn::normalized_sigma_2_plus_infty}, $\sigma_\text{max}/|R|<0$ when $R$ is sufficiently large, hence there exists a solution to the critical reaction rate for stability criteria A and B ($\sigma_\text{max}=0$ and $\sigma_\text{max}/|R|=1$) when $R>0$.

Substituting the free energy model (Eq.~\ref{eqn::regular_solution}) and $\gamma_\dagger$ into Eq.~\ref{eqn::normalized_sigma_2_minus_infty},
we have $\lim_{R\to-\infty}{\sigma_\text{max}/|R|}=c^{-1}+2\Omega$. Appendix \ref{appendix:limit} proves that this quantity is greater than 1 in the spinodal region.
Therefore, when $R<0$, stability criteria A and B cannot be satisfied in the spinodal region.

Fig.~\ref{fig::no_stress} shows the concentration field on the $\text{Da}_0^{-1}-\bar{R}/k_0$ phase diagram ($\bar{R}>0$), with snapshots taken at an average concentration of 0.3, 0.5 and 0.7. We observe that phase separation is suppressed at high reaction rates and low diffusivity, where the concentration field is uniform.
At low reaction rates and high diffusivity, the patterns coarsen more.

The boundary between stability and phase separation follows two regimes: At low diffusivity, the critical reaction rate is independent of diffusivity. At higher diffusivity, the critical reaction rate $R_\text{crit}$ depends linearly on relative diffusivity $\text{Da}_0^{-1}$, in agreement with the scaling mentioned in sec.~\ref{sec::reaction_kinetics}.
In the latter regime, the competition between reaction and diffusion leads to patterns that resemble spinodal decomposition, in agreement with a nonzero most unstable wavenumber (Eq.~\ref{eqn::k_max}).

The scaling of the critical reaction rate is well estimated analytically by the stability criteria, which are the region above the stability boundaries indicated by the solid curves.
Among the three stability criteria studied, criterion C ($0<c_0(t=0)+\sigma_0 e^{K}<1$) agrees best with the region of suppressed phase separation obtained from numerical simulations, offering significant improvement in accuracy compared to $\sigma_\text{max}/|R|<1$ and $\sigma_\text{max}<0$.

SI Movie 1 shows the temporal evolution of the simulations and all three stability criteria. Stability criterion C agrees with the numerical simulation at all times as the average concentration increases from 0.1 to 0.9. As the average concentration increases from 0.1 to 0.9, the critical reaction rate at a given $\text{Da}_0$ increases, which corresponds to a shrinking stability region in the $\text{Da}_0^{-1}-\bar{R}/k_0$ phase diagram whose boundary shifts upward. In contrast, criteria A and B, which are only determined by the average concentration at each point in time, predict that the critical reaction rate increases and then decreases. A large discrepancy between criteria A/B and simulations exists when $c_0$ is close to 0.1 or 0.9. This highlights the importance of history dependence in pattern formation in an open and chemically-driven system.

\begin{figure*}[htb]
  \includegraphics[width=\columnwidth]{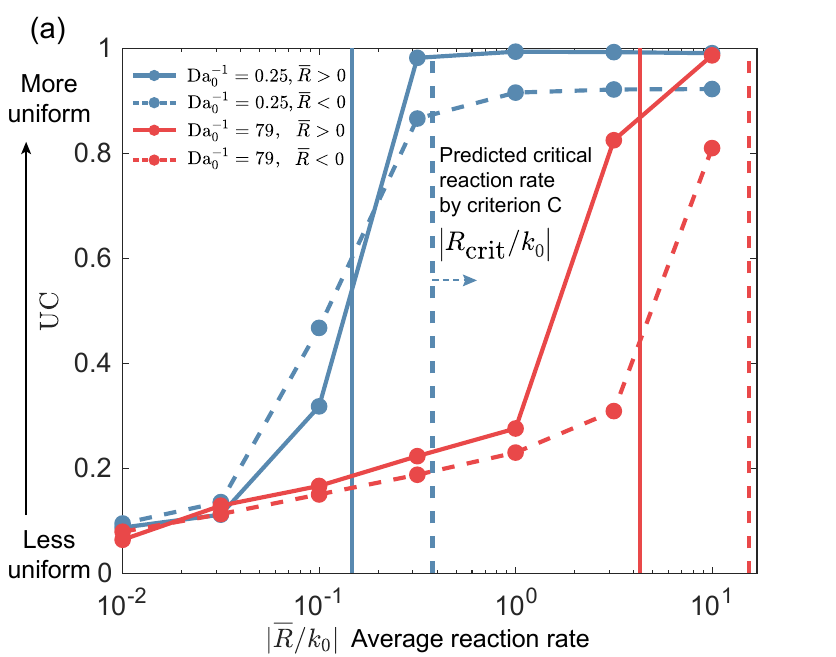}
  \includegraphics[width=0.95\columnwidth]{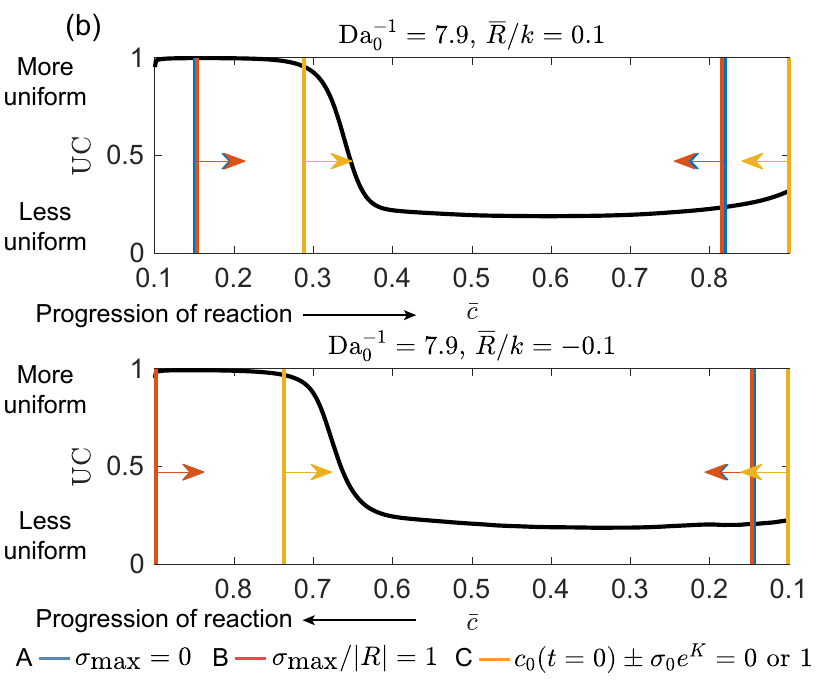}
  \caption{(a) The uniformity coefficient (UC) of the concentration field as a function of normalized reaction rate $|\bar{R}/k_0|$ at $\text{Da}_0^{-1}=0.25$ (blue) and $79$ (red) at the average fraction at 0.5, which correspond to snapshots in the first and sixth columns in Fig.~\ref{fig::no_stress}(b) (solid curves, $\text{R}>0$) and Fig. S1(b) (dashed curves, $\bar{R}<0$), respectively. The vertical lines are the normalized critical reaction rate $|R_\text{crit}/k_0|$ obtained using criterion C, above which phase separation is suppressed and hence a high uniform coefficient is expected. Solid and dashed curves correspond to $\bar{R}>0$ and $\bar{R}<0$, respectively. (b) The temporal evolution of UC (as a function of average concentration $\bar{c}$) at $\text{Da}_0=7.9$ and $\bar{R}/k=\pm 0.1$ is shown as black curves. The analytical stability criteria (blue, red, and yellow refer to criteria A, B, and C, respectively) predict that phase separation (and hence low uniformity coefficient) occurs between the two vertical lines. }
  \label{fig::no_stress_UC}
\end{figure*}

Fig. S1 and SI movie 2 show the comparison of simulations and stability boundaries when the average reaction rate is negative $\bar{R}<0$, as the system starts backward from a uniform concentration field of $c_0=0.9$ with added noise.
As explained above, the boundaries of stability criteria A and B do not exist when in the spinodal region and $R<0$. However, we observe from numerical simulations that phase separation can be suppressed at high reaction rates $|\bar{R}|$ and low diffusivity, and that the critical reaction rate $|R_\text{crit}|$ at a given $\text{Da}_0^{-1}$ increases over time. Compared to forward reaction, backward reaction induces a greater extent of heterogeneity in the concentration field at the same extent of reaction, as expected from the asymmetry in the normalized maximum instability rate (Eqs.~\ref{eqn::normalized_sigma_2_plus_infty} and \ref{eqn::normalized_sigma_2_minus_infty}). Here, the asymmetry results from the asymmetry of the prefactor $R_0$. Physically, because the reaction requires the presence of vacancy, the reaction slows down as $c$ increases, hence the forward reaction is autoinhibitory while the backward reaction is autocatalytic, which results in stronger heterogeneity in the latter.
Correspondingly, the stability boundary from criterion C also shows a higher critical reaction rate $|R_\text{crit}|$ at the same extent of reaction and $\text{Da}_0^{-1}$.
While criteria A and B are nonexistent for backward reaction in Fig. S1, criterion C agrees well with the simulations, again highlighting the importance of considering the dependence of $\sigma$ on $c_0$.

To quantify the level of phase separation based on the simulations, we define a uniformity coefficient UC that measures the uniformity of the concentration field
\begin{equation}
  \text{UC} = 1 - \sqrt{\frac{\int{(c-\bar{c})^2 d\mathbf{x}}}{\bar{c}(1-\bar{c})\int{d\mathbf{x}}}}
\end{equation}
where $\bar{c}$ is the average concentration. In other words, $1-\text{UC}$ is the standard deviation of $c(\mathbf{x})$ normalized by the maximum possible standard deviation at a given $\bar{c}$. $\text{UC}=1$ when $c(\mathbf{x})$ is uniform; $\text{UC}=0$ when $c(\mathbf{x})$ is equal to either 0 and 1 (fully phase separated).

Fig.~\ref{fig::no_stress_UC}(a) summarizes UC values for some simulations in Fig.~\ref{fig::no_stress} and Fig. S1 and confirms that at a given $\text{Da}_0$, UC increases with increasing reaction rates, that is, the concentration field becomes more uniform, and that smaller $\text{Da}_0^{-1}$ (smaller diffusivity) results in increased uniformity. The UC values are taken at $\bar{c}=0.5$, corresponding to the same extent of reaction for both forward and backward reactions. For most data points, UC is higher for forward compared to backward reactions.
The critical reaction rate based on criterion C which is represented by the vertical lines is located in the rapidly increasing region of the $\text{UC}-\log{|\bar{R}/k_0|}$ curve.

Fig.~\ref{fig::no_stress_UC}(b) shows the temporal evolution of UC under constant total reaction rate at $\text{Da}_0=7.9$ and $\bar{R}/k=\pm 0.1$. The uniformity remains high initially until it quickly decreases to a lower value, indicating the onset of the phase separation. The onset is earlier for backward reaction than for forward reaction at the same rate. The stability criteria predict that phase separation occurs in the region between the two lines. Criteria A and B predict a much earlier onset and earlier termination of phase separation than the simulation. In contrast, criterion C predicts the onset with improved accuracy and correctly predicts that phase separation persists all the way to the end of the reaction.

In summary, Fig.~\ref{fig::no_stress_UC} recapitulates the observations that criterion C provides a good estimate of the condition for the suppression of phase separation and the asymmetry between forward and backward reactions.

\section{LFP nanoparticles}
\label{sec::chemomechanics}
In this section, we apply the theory to a specific system: lithium iron phosphate (LFP) nanoparticles undergoing Li insertion and de-insertion. We consider the chemo-mechanical coupling in the system and study the effect of reaction rate and lateral diffusivity on the suppression of phase separation.

LFP is an important energy storage material that is widely used in Li-ion batteries\cite{Padhi1997,Kang2009}.
At equilibrium, LFP can phase separates into a Li-poor phase $\text{Li}_\delta \text{FePO}_4$ and a Li-rich phase $\text{Li}_{1-\delta}\text{FePO}_4$ ($\delta=0.035$ at room temperature)\cite{Bai2011,Cogswell2012,Delmas2008,Yamada2006}.

As the cathode in the battery, lithium ions in the electrolyte are inserted into iron phosphate crystal during discharge: $\text{Li}^+ + e^- + \text{FePO}_4 \to \text{LiFePO}_4$. The opposite reaction occurs during charge.
Li is inserted into the crystal from the (010) plane (or a-c plane) as shown in the schematics in Fig.~\ref{fig::schematics}. Diffusion in the (010) plane can occur due to defects\cite{malik2010} or at the solid-liquid interface via fluid-enhanced surface diffusion\cite{Li2018a}. On the other hand, due to the crystal anisotropy, diffusion in the [010] direction is fast \cite{malik2010}. As a result, thin platelet particles can be described by a 2D model in the (010) facet by depth averaging in the [010] direction\cite{Nadkarni2018}.

Previous work has used the Allen-Cahn reaction model which corresponds to the model in the work without diffusion to simulate the dynamics of phase transformation in LFP particles undergoing charge and discharge\cite{Bai2011,Cogswell2012}, the effect of surface wetting and nucleation\cite{Cogswell2013}, and size dependence\cite{Cogswell2018a}. Recently this model has been used to extract the free energy, reaction kinetics, and surface heterogeneity from experimental images\cite{Zhao2023LFP}.

Ref. \cite{Li2018a} revealed that when particles are immersed in electrolytes compared to no electrolyte exposure, the organic solvent or water can enhance surface diffusion and promote phase separation. Recently, it has been shown that the effect of electro-autocatalysis can lead to dynamic surface phases.
Inspired by these experimental results, in this work, we study the interplay between reaction and diffusion and the resulting pattern formation during Li insertion and de-insertion, at different reaction rates (which correspond to charge/discharge rates) and lateral diffusivity in the (010) plane.

In crystalline solids, an intercalation reaction such as in LFP $\text{M}_\text{res} \to \text{M}$ can often cause deformation of the crystal lattice. A heterogeneous concentration of M due to phase separation can lead to the local accumulation of stress, which in turn affects the phase separation. Therefore, the model introduced in sec.~\ref{sec::model} needs to be generalized to account for the chemo-mechanical coupling.

Here, we consider a solid crystal that remains coherent as the reaction takes place, that is, the displacement field $\mathbf{u}$ relative to the configuration when the concentration of M is uniform prior to the reaction varies continuously in space and the crystal sustains an elastic strain $\boldsymbol\varepsilon^{\text{el}}$ to accommodate coexisting phases (or heterogeneous concentration field) that have different lattice constants.

A model of coherent phase separation in solids can be formulated based on the Cahn-Hilliard model and the theory of inclusion\cite{khachaturyan2013theory,Cahn1984}. The total free energy $G$ consists of chemical free energy (Eq.~\ref{eqn::chemical_free_energy}) and mechanical energy. In the small deformation regime,
\begin{equation} \label{eqn::bulk_free_energy}
  G=\int_\Omega {\left( c_s \left( g_h(c) + \frac{1}{2} \kappa | \nabla c|^2  \right) + \frac{1}{2} \lambda_{ijkl} \varepsilon_{ij}^{\text{el}} \varepsilon_{kl}^{\text{el}} \right) dV }
\end{equation}
where $c_s$ is the lattice site density, $c$ is the fraction of lattice sites occupied by M, and  $\lambda_{ijkl}$ is the stiffness tensor.
Suppose the lattice constants under stress-free conditions depend linearly on the occupied fraction $c$ (known as Vegard's law \cite{Ashcroft1991}), $\varepsilon_{ij}^0 c$, then the elastic strain is
\begin{equation}
  \varepsilon_{ij}^{\text{el}} = \frac{1}{2} \left( \partial_j u_i + \partial_i u_j \right) - \varepsilon_{ij}^0 c
\end{equation}
The mechanical equilibrium condition can be derived from $\delta G / \delta \mathbf{u}=0$:
\begin{equation} \label{eqn::mech_eqm}
\nabla \cdot \bm{\sigma} = 0
\end{equation}
where the stress tensor is (using Einstein notation)
\begin{equation} \label{eqn::stress}
\sigma_{ij} = \lambda_{ijkl}\varepsilon_{ij}^{\text{el}}
\end{equation}
Because mechanical relaxation is much faster than reaction and diffusion, we assume mechanical equilibrium (Eq.~\ref{eqn::mech_eqm}) holds at all times.
Similar to Eq.~\ref{eqn::chemical_potential}, the chemical potential is defined by the variational derivative
\begin{equation} \label{eqn::mu}
  \mu = \mu_h(c) - \kappa \nabla^2 c - c_s^{-1} \varepsilon_{ij}^0 \sigma_{ij}
\end{equation}
One can solve for the displacement field analytically in the Fourier space in an infinitely large domain \cite{khachaturyan2013theory,Cogswell2012} and substitute it into the total free energy to obtain
\begin{multline}
  G =  \int_V { c_s \left( g_h(c) + \frac{1}{2} \kappa | \nabla c|^2  \right)  d\mathbf{r}} \\ + \frac{1}{2 \left( 2 \pi \right)^d} \int_{\hat{V}}{ B(\mathbf{n}) \left[ \Delta \hat{c}(\mathbf{k}) \right]^2 d \mathbf{k}}
\end{multline}
where $d$ is the dimensionality, $\Delta\hat{c}$ is the Fourier transform of $c-\bar{c}$, the deviation of concentration field from the mean, and
\begin{equation}
  B(\mathbf{n}) = \lambda_{ijkl}\varepsilon_{ij}^0\varepsilon_{kl}^0 - n_i \sigma_{ij}^0 \Omega_{jl}(\mathbf{n}) \sigma_{lm}^0 n_m
\end{equation}
where $\mathbf{n} = \mathbf{k} /|\mathbf{k}|$ is the unit vector in Fourier space, and
\begin{equation}
  \left(\Omega^{-1}\right)_{ik} = \lambda_{ijkl}n_j n_l
\end{equation}
The expression of mechanical energy by an integral in Fourier space corresponds to an additional term in the instability growth rate of the concentration field
\begin{equation}
  \sigma = \pderiv{R}{c} - \left( -\pderiv{R}{\tilde{\mu}} + L |\mathbf{k}|^2 \right) \left( \mu_h' + \kappa |\mathbf{k}|^2 + B(\mathbf{n}) \right)
\end{equation}
For an isotropic material, $B(\mathbf{n})$ is a constant, hence instability develops isotropically, all the analysis in section \ref{sec::linear_stability} holds simply by absorbing $B(\mathbf{n})$ into $\mu_h'$. For anisotropic solids, the maximum instability growth rate $\max_\mathbf{n}{\sigma}$ is obtained at $\mathbf{n}^* = \text{argmin}B(\mathbf{n})$, the instability develops preferentially along the direction of $\mathbf{n}^*$ in Fourier space.
Therefore, the maximum instability rate $\sigma_\text{max}$ and the most unstable wavenumber $k_\text{max}$ are obtained along $\mathbf{n}^*$ and can be calculated using Eq.~\ref{eqn::sigma_max} and Eq.~\ref{eqn::k_max} by substituting $\mu_h'$ with $\mu_h' + B(\mathbf{n}^*)$.

\begin{figure*}
  \includegraphics[width=\textwidth]{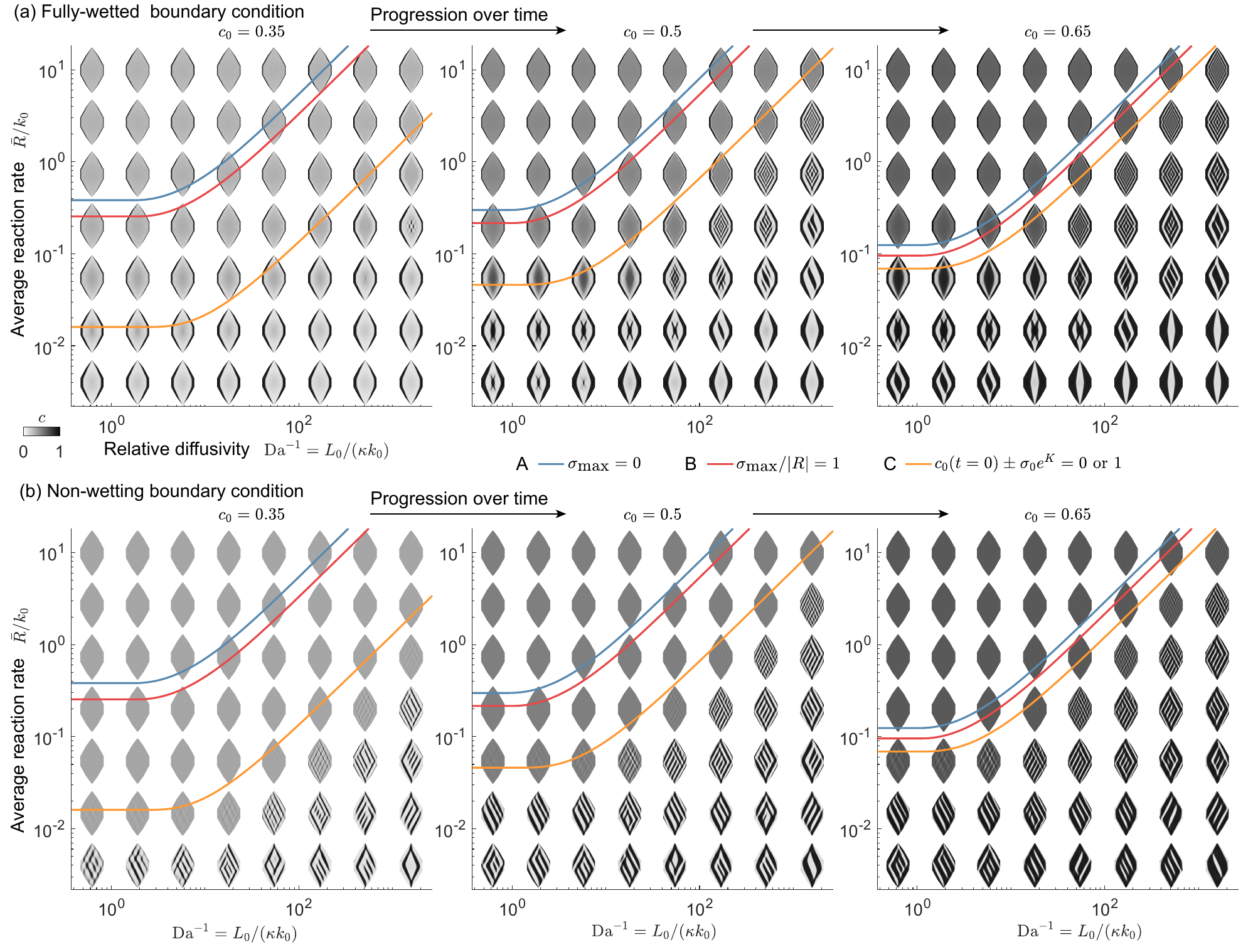}
  \caption{Phase transformation in LFP particle during Li insertion in the $\text{Da}_0^{-1}-\bar{R}/k_0$ phase diagram using (a) fully-wetted boundary condition and (b) non-wetting boundary condition.  Simulations start from an average fraction of 0.2. Constant average reaction rate $\bar{R}$ is imposed. From left to right, the three panels show snapshots when the average concentrations are 0.35, 0.5, and 0.75, respectively. The colored solid curves denote the boundaries of the analytical stability criteria A-C.}
  \label{fig::with_stress}
\end{figure*}

We apply the theory to a C3-shaped LFP crystal particle\cite{Smith2012} with thermodynamic and mechanical parameters given in the literature \cite{Maxisch2006,Cogswell2012,Deng2022} ($\Omega=4.48 k_B T$, $\sqrt{\kappa/(k_B T)}=3\text{nm}$, $B=0.19\text{GPa}$, or $B/c_s=3.36 k_B T$, values for $\lambda_{ijkl}$ and $\epsilon_{ij}^0$ can be found in Ref.~\cite{Zhao2023LFP}).
In LFP, the insertion of lithium causes an expansion in the [100] direction and contraction in the [001] direction. As a result, the interface between FP and LFP prefers to be aligned along the [101] direction, as shown in Fig.~\ref{fig::schematics}.
It has been shown that in finite-sized particles, the phase-separated domains form stripes in the [101] direction ($\mathbf{n}^*)$ of a certain wavelength as the system minimizes the interfacial and elastic energy of inclusions \cite{khachaturyan2013theory,Cogswell2012,Cogswell2018a}. Here, we choose a particle size that is $500\text{nm}$ along the [001] direction, which exceeds the spacing between stripes \cite{Cogswell2018a}. As a result, multiple stripes can form within the particle.

Following Ref. \cite{Cogswell2012}, we impose the mechanical boundary condition that the heterogeneous displacement field ($\varepsilon_{ij}-\varepsilon_{ij}^0\bar{c}$) is zero \cite{khachaturyan2013theory} and no flux boundary condition $\mathbf{n}\cdot \nabla \mu=0$ since intercalation reaction cannot occur at the side facets.
We consider two types of wetting boundary conditions. The first one is where Li fully wets the side facets of the platelet particle\cite{Cogswell2013,Cogswell2018a}. Such surface wetting can play an important role in the nucleation of phases. We set $c=0.99$ on all the boundaries in the 2D simulations.
When using the fully wetted boundary condition, the initial condition is fully equilibrated at zero reaction rate, at the desired average fraction, and with the given boundary condition.
The second boundary condition is the non-wetting boundary condition used in section \ref{sec::simulation} ($\mathbf{n} \cdot \nabla c=0$).
In this case, the simulations start from a uniform concentration field with additive white noise ($\sigma_0=0.02$).
For all cases, the particles are subject to the constraint of a constant average reaction rate. For Li insertion, the average fraction goes from 0.2 to 0.8, and the opposite for Li de-insertion.
The same reaction kinetics is used for LFP (Eqs.~\ref{eqn::R_c_Delta_mu} and \ref{eqn::generalized_R0}) in this section.

\begin{figure*}
  \includegraphics[width=\columnwidth]{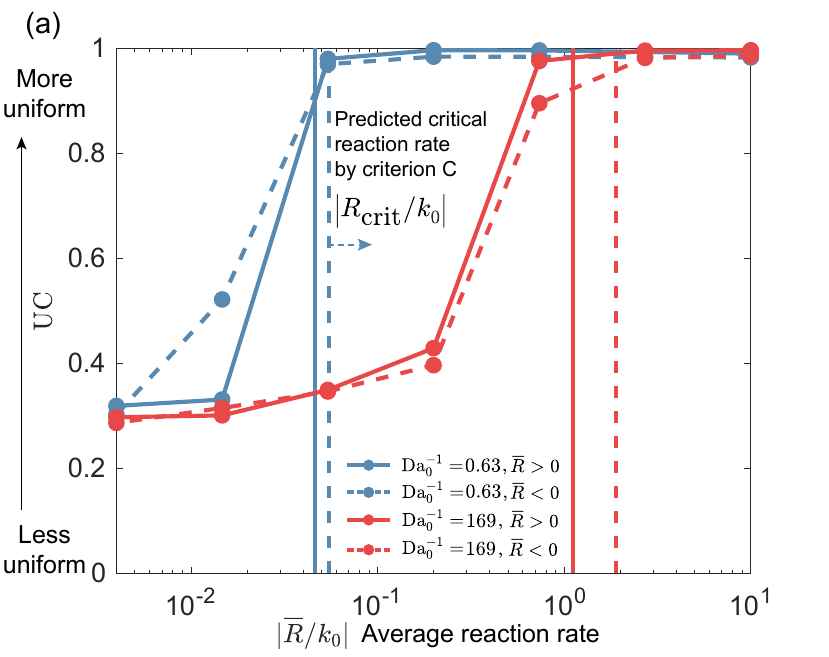}
  \includegraphics[width=0.9\columnwidth]{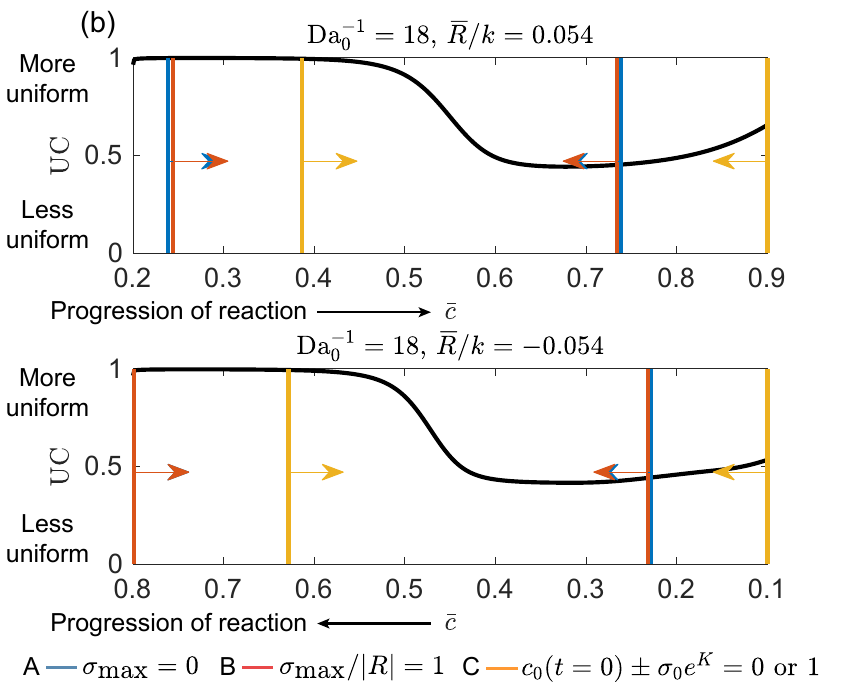}
  \caption{(a) UC of LFP particle as a function of normalized reaction rate $|R/k_0|$ at $\text{Da}_0^{-1}=0.63$ (blue) and $169$ (red) at the average fraction at 0.5, which correspond to snapshots in the first and sixth columns in Fig.~\ref{fig::with_stress} (solid curves, $\text{R}>0$) and Fig. S2 (dashed curves, $\text{R}<0$). The vertical lines are the normalized critical reaction rate $\text{R}_\text{crit}/k_0$ obtained using criterion C. Solid and curves correspond to $R>0$ and $R<0$, respectively. (b) The temporal evolution of UC at $\text{Da}_0=18$ and $R/k=\pm 0.054$ is shown as black curves. The analytical stability criteria predict that phase separation occurs in between two vertical lines. }
  \label{fig::with_stress_UC}
\end{figure*}

Fig.~\ref{fig::with_stress} shows the simulations in the phase diagram of $\text{Da}_0^{-1}-\bar{R}/k_0$.
Wetting of Li at the boundaries, as shown in Fig.~\ref{fig::with_stress}, promotes nucleation and hence the growth of the Li-rich phase at the boundary. When particles are small enough, the phase transformation is dominated by surface wetting\cite{Cogswell2018a}. In our case, however, because the particles are large enough that multiple stripes can form, the morphology is determined by the interplay between surface wetting and bulk elasticity.
As the average fraction increases, diamond-shaped patterns or stripes form in the bulk. Due to the surface wetting, the orientation of the Li-poor and Li-rich interface is not always along [101] direction. In contrast, without the constraint of the surface wetting as applied in Fig.~\ref{fig::with_stress}(b), the morphology becomes substantially different, as most stripes are now aligned in the energetically favorable [101] direction.

For both boundary conditions, phase separation can be suppressed in the bulk at a high reaction rate and low diffusivity. On the other end of the phase diagram with a low reaction rate and high reaction where the system is closer to equilibrium, the patterns coarsen and the interfacial area and the number of stripes become smaller.

Similar to Fig.~\ref{fig::no_stress}, the critical reaction rate is independent of $\text{Da}_0$ at low diffusivity while it scales linearly with $\text{Da}_0^{-1}$ at higher diffusivity. In the regime of linear scaling and close to the boundary, checkerboard patterns aligned in the [101] directions that resemble spinodal decomposition emerge as a result of nonzero most unstable wavenumber $k_\text{max}$ in the most unstable direction $\mathbf{n}^*$.

The estimated critical reaction rates based on the analytical stability criteria are shown as solid curves in Fig.~\ref{fig::with_stress}. Notice that under the fully-wetted boundary condition, the concentration field is never uniform, hence the stability criteria we derived from linear stability analysis do not apply. Nevertheless, we superimpose the curves of critical reaction rate from Fig.~\ref{fig::with_stress}b on the simulations in Fig.~\ref{fig::with_stress}a, and find that these analytical results (especially criterion C) are still fairly good estimates of the boundary of phase separation.
Again, the scaling of the critical reaction rate is captured by the stability criteria, and criterion C gives a much more accurate estimate than criteria A and B throughout the entire process, as confirmed by SI movies 3 and 4 which show the temporal evolution of the simulations and stability criteria for both boundary conditions.

Fig. S2(a-b) and SI movie 5-6 show the comparison of simulations and stability boundary using both boundary conditions when the average reaction rate $\bar{R}<0$. The average fraction at the initial condition is 0.8. Similar to sec.~\ref{sec::simulation}, phase separation can be suppressed at high reaction rates $|R|$ and low relative diffusivity, and the stability boundary predicted by criterion C agrees well with simulation, while criteria A and B do not exist for Li de-insertion in the spinodal region.

Since the same reaction kinetics is used as in sec.~\ref{sec::simulation}, compared to forward reaction, backward reaction also leads to a greater level of heterogeneity at the same extent of reaction. 
When the fully-wetted boundary condition is used, there always exists a layer of Li-rich phase at the boundaries, as a result, the asymmetry in the morphology of phase separation between Li insertion and de-insertion is more pronounced compared to the non-wetting boundary condition, consistent with the findings in Ref. \cite{Cogswell2018a}.



In Fig.~\ref{fig::with_stress_UC}(a), we summarize the observations above quantitatively by showing the uniformity coefficient at certain values of $\text{Da}_0$ and $R/k_0$ and compare with the critical reaction rate predicted by stability criteria. We see that the uniformity increases with increasing reaction rates and decreasing diffusivity. The critical reaction rate based on criterion C (vertical lines) is a good estimate of the condition for the suppression of phase separation. Fig.~\ref{fig::with_stress_UC}(b) shows the evolution of UC over time. Similar to Fig.~\ref{fig::no_stress_UC}, criterion C provides an estimate for the duration of phase separation that agrees with the simulation.


\section{Discussion}
In this work, we studied pattern formation in chemically driven and open bulk systems using a reaction-diffusion model for phase-separating systems. We find that reaction rate and diffusivity strongly affect the patterns and the competition between the two determines the morphology -- high reaction rate and low diffusivity suppress phase separation, low reaction rate and high diffusivity allow the patterns to coarsen, while spinodal patterns emerge between the two limits.

The critical reaction rate for the suppression of phase separation depends on the specific reaction kinetics of the system as well as the direction of the reaction. A concentration-dependent kinetic prefactor $R_0$ can alter the stability of the system when chemically driven -- an autocatalytic reaction enhances phase separation while an autoinhibitory reaction suppresses phase separation.

Numerical simulations performed at constant average reaction rate confirm the scaling law of the critical reaction rate $R_\text{crit}$ with respect to Damkholer number $\text{Da}_0$ derived from linear stability analysis: $R_\text{crit}$ is independent of $\text{Da}_0$ when the system is reaction-limited, while $R_\text{crit} \propto \text{Da}_0^{-1}$ when the system is diffusion-limited. Near the stability boundary and in the diffusion-limited regime, patterns of a certain wavelength emerge, resembling spinodal decomposition in non-driven systems.

We found a quantitative stability criterion that takes into account the concentration-dependent instability growth rate over the course of the reaction by integrating the time-dependent instability growth rate and following the growth of perturbation in time, therefore providing a more accurate estimate for when phase separation occurs throughout the simulation compared to the instantaneous instability growth rate, which highlights the importance of history dependence in pattern formation in chemically driven systems.

Coupling the reaction-diffusion model with mechanical deformation, we applied the theory to lithium iron phosphate (LFP) nanoparticles.
The chemo-mechanical coupling and surface wetting have a significant influence on pattern formation in LFP. Stripes and diamond-shaped patterns form when phase separation occurs. The critical reaction rate that suppresses phase separation in LFP particles follows the same scaling as the general phase-separating system with isotropic properties. Because phase separation within particles leads to the accumulation of stress in order to accommodate phases with different lattice parameters and eventually causes mechanical failure, the result suggests that reducing lateral diffusion in LFP by engineering the particle surface to lower surface diffusion or optimizing synthesis to reduce defect-assisted diffusion in the bulk may improve its performance. The suppression of phase separation at high reaction rates also explains the high rate capability of LFP despite being a phase separation material \cite{Padhi1997,Kang2009,Bazant2013}.
A recent study of the evolution of lithium composition in [100]-oriented LFP nanoparticles observed dynamic surface phase separation\cite{Koo2023}, further motivating the extension of the theory to study the pattern formation in 3D at varying reaction rates and diffusivity.

Recently, X-ray images of LFP nanoparticles during lithium insertion and de-insertion have been used to learn the heterogeneous reaction kinetics pixel-by-pixel based on the Allen-Cahn reaction model\cite{Zhao2023LFP}. Our analysis of both diffusion and reaction in this work suggests that the image inversion method may be further generalized to learn the dynamics of diffusion, together with images from other crystallographic planes such as the [100]-oriented LFP nanoparticles mentioned above. 

In this work, we used Butler-Volmer kinetics to model electrochemical reactions. Ref.~\cite{Fraggedakis2020a} also analyzed the stability when the electrochemical reaction is described by coupled ion-electron transfer kinetics (CIET) \cite{Fraggedakis2021}, which considers the simultaneous transfer of ion and electron, and revealed the importance of reaction kinetics in controlling interfacial stability. With a growing amount of experimental evidence that suggests that CIET theory captures the kinetics of lithium intercalation in LFP \cite{Zhao2023LFP} and lithium transition metal oxides, elucidating the effect of CIET on pattern formation in lithium intercalation materials and extending the analysis to more generalized reaction kinetics such as a recently developed quantum theory that unifies Butler-Volmer and Marcus kinetics \cite{Bazant2023} will be an important future direction and crucial in understanding and optimizing the rate capability of energy storage materials.

While we focused on systems that phase separate at thermodynamic equilibrium, the theory can also be extended to study intra-particle ``fictitious phase separation'', a phenomenon that is observed in non-phase-separating lithium transition metal oxides during delithiation (lithium de-insertion) that otherwise does not phase separate at equilibrium or during lithiation \cite{Park2021,Zhou2016,Grenier2017,Marker2019}. The fictitious phase separation has been attributed to the destabilizing electro-autocatalytic effect during delithiation that leads to a bimodal distribution of a population of lithium-rich and lithium-poor particles within the porous electrode \cite{Park2021}.
Motivated by recent operando imaging of lithium composition in lithium transition metal oxides\cite{Merryweather2021,Xu2022}, it will be useful to understand the effect of reaction kinetics and diffusion on pattern formation in these systems.

More broadly, the theory developed in this work can be applied to other chemically driven systems such as electrochemical reactions and biological systems. For example, the theory can be applied to study the pattern formation in hydrogen absorption in palladium\cite{Narayan2017,Bardhan2013} and electrooxidation of $\text{CO}_2$ on platinum. Our analysis can be extended to study the adsorption, oxidation, diffusion, and pattern formation of $\text{CO}_2$ on the platinum surface when current or voltage is controlled\cite{Krischer2000,Varela2005}, which is known to undergo oscillations and exhibit traveling waves on the electrode\cite{Kim2001,Bonnefont2005}.

The theory may also be applicable to pattern formation and phase separation in biological systems such as the Min system responsible for cell division, which consists of MinD, MinC, and MinE proteins and undergoes pole-to-pole oscillations in \emph{E. coli} cells and is known to also give rise to surface waves and spirals in vitro \cite{Loose2008,}. In addition, studying the effect of ATP-driven reactions on liquid-liquid phase separation in biological cells (also known as biomolecular condensates)\cite{Brangwynne2015,Shin2017,Berry2018} using a thermodynamically consistent reaction-diffusion model \cite{Kirschbaum2021,Zwicker2022} will also be an interesting future direction.

\bibliography{Zhao}

\appendix
\section{Normalized instability rate}
\label{appendix::norm_instability}
Here we show that if $R_0$ only depends on $c$, $g(\Delta \tilde{\mu})$ follows Eq.~\ref{eqn::BV}, and $\mu_h'<0$, then $\sigma_\text{max}/|R|$ decreases with increasing $|R|$.
$|\partial_{\tilde{\mu}} R|$ increasing with increasing $|\Delta \tilde{\mu}|$ and hence increasing reaction rate $|R|$, hence $\text{Da}$ increases and $F(\text{Da})$ decreases. At the same time, $g'/|g|$ decreases,
because $g(x)=e^{\alpha x}-e^{-(1-\alpha)x}$, $g'(x) = \alpha e^{\alpha x}+(1-\alpha)e^{-(1-\alpha)x}>0$,
\begin{equation}
  \left( \frac{g'}{g} \right)' = \frac{gg'' -g^{\prime2}}{g^2} = - \frac{1}{g^2} (2\alpha-1)^2 e^{(2\alpha-1)x}  \leq 0
\end{equation}
therefore $g'/|g|$ decreases with increasing $|x|$ and hence increasing $|R|$.
Finally, because when $R_0$ only depends on $c$,
\begin{equation}
  \frac{\sigma_\text{max}}{|R|} = \text{sgn}(R) \pderiv{\ln{R_0}}{c} - \frac{g'}{|g|} \mu_h' F(\text{Da})
\end{equation}
we find that $\sigma_\text{max}/|R|$ decreases with increasing reaction rate $|R|$. So the minimum is achieved at infinite $|R|$.
Because $g'/|g| \to \alpha$ as $\Delta \tilde{\mu} \to \infty$ and $g'/|g| \to 1-\alpha$ as $\Delta \tilde{\mu} \to -\infty$, we obtain the expressions for $\lim_{R\to \pm \infty}{\sigma_\text{max}/|R|}$ in Eqs.~\ref{eqn::normalized_sigma_BV_plus_infty} and \ref{eqn::normalized_sigma_BV_minus_infty}.

If $R_0$ only depends on $c$, and Butler-Volmer kinetics is symmetric, that is $\alpha=0.5$, then we have $g(x)=2\sinh{(x/2)}$, $g'(x)=\cosh{(x/2)}$, $g^{\prime2} - (g/2)^2 = 1$.
Hence the normalized maximum instability growth rate can be expressed explicitly in terms of $R$,
\begin{equation} \label{eqn::normalized_sigma_BV}
  \frac{\sigma_\text{max}}{R} =  \pderiv{\ln{R_0}}{c}  -  \text{sgn}(R) \sqrt{\frac{1}{4}+\left(\frac{R_0}{R}\right)^2} \mu_h' F(\text{Da})
\end{equation}


Using the generalized Butler-Volmer kinetics (Eq.~\ref{eqn::generalized_R0} and Eq.~\ref{eqn::BV}), the reaction rate can be written as
\begin{equation}
  R = \frac{k_0}{\gamma_\dagger} \left[ e^{\alpha\mu_\text{res}}-e^{-(1-\alpha)\mu_\text{res}+\mu} \right]
\end{equation}
Since
\begin{equation} \label{eqn::dh_h}
  \pderiv{\ln{R_0}}{\mu}  = \frac{1}{1-e^{\Delta \tilde{\mu}}} = \frac{1}{2} \left( 1- \text{coth}\frac{\Delta \tilde{\mu}}{2} \right)
\end{equation}
When $\alpha=0.5$, $R=2 k_0 \gamma_\dagger^{-1} e^{\alpha\tilde{\mu}}\sinh{\Delta \tilde{\mu}/2}=2R_0\sinh{\Delta \tilde{\mu}/2}$,
Eq.~\ref{eqn::dh_h} can also be written in a form that eliminates the dependence on the affinity $\Delta \tilde{\mu}$,
\begin{equation} \label{eqn::dh_h_alt}
  \pderiv{\ln{R_0}}{c}  = \frac{1}{2} - \text{sgn}(R) \sqrt{\frac{1}{4}+\left(\frac{R_0}{R}\right)^2 }
\end{equation}
This can also be obtained using $\partial_\mu \ln{R} = \partial_\mu \ln{R_0} - g'/g$ and noting that $ \partial_\mu \ln{R_0} =\alpha$.
Hence we obtain a form of Eq.~\ref{eqn::normalized_sigma} that depends on $R$ explicitly,
\begin{equation} \label{eqn::normalized_sigma_generalized_BV}
  \frac{\sigma_\text{max}}{R} =  \pderiv{\ln{R_0}}{c} + \left[ \frac{1}{2} -  \text{sgn}(R) \sqrt{\frac{1}{4}+\left(\frac{R_0}{R}\right)^2} \right] \mu_h' F(\text{Da})
\end{equation}

\section{Limit of normalized instability rate}
\label{appendix:limit}
Using the regular solution model as the free energy model and $\gamma_\dagger=1-c$, based on Eq.~\ref{eqn::normalized_sigma_2_minus_infty},
we now prove that $X \equiv \lim_{R\to-\infty}{\sigma_\text{max}/|R|}=-c^{-1}+2\Omega>1$.
Within the spinodal region, $c>c_- = (1-\sqrt{1-2/\Omega})/2$. Therefore
\begin{equation}
  X > 2\Omega - \frac{2}{1-\sqrt{1-\frac{2}{\Omega}}} \equiv h(\Omega)
\end{equation}
When $\Omega>2$, the right hand side decreases with increasing $\Omega$,
Because $1-\sqrt{1-\frac{2}{\Omega}}=\Omega^{-1}+\Omega^{-2}/2+o(\Omega^{-2})$, we obtain that $\lim_{\Omega\to\infty}{h(\Omega)}=1$. Therefore, $X>1$.

\end{document}